\pdfoutput=1
\documentclass[aps,twocolumn,superscriptaddress,letterpaper]{revtex4}
\usepackage{graphicx,amssymb,bm,natbib,amsfonts,amsmath}
\usepackage{hyperref}

\newcommand{\kpar} {$k_\parallel$}
\newcommand{\kz}{$k_z$}

\begin{document}

\title{Departure from the conical dispersion in epitaxial graphene}

\author{S.Y. Zhou}
\affiliation{Department of Physics, University of California,
Berkeley, CA 94720, USA}
\affiliation{Materials Sciences Division,
Lawrence Berkeley National Laboratory, Berkeley, CA 94720, USA}
\author{D.A. Siegel}
\affiliation{Department of Physics, University of California,
Berkeley, CA 94720, USA}
\affiliation{Materials Sciences Division,
Lawrence Berkeley National Laboratory, Berkeley, CA 94720, USA}
\author{A.V.~Fedorov}
\affiliation{Advanced Light Source, Lawrence Berkeley National
Laboratory, Berkeley, California 94720, USA}
\author{A. Lanzara}
\affiliation{Department of Physics, University of California,
Berkeley, CA 94720, USA}
\affiliation{Materials Sciences Division,
Lawrence Berkeley National Laboratory, Berkeley, CA 94720, USA}

\date{\today}


\begin{abstract}

The $\pi$ bands of epitaxially grown graphene are studied by using high resolution angle resolved photoemission spectroscopy.  Clear deviations from the conical dispersion expected for massless Dirac fermions and an anomalous increase of the scattering rate are observed in the vicinity of the Dirac point energy.  Possible explanations for such anomalies are discussed in terms of many-body interactions and the opening of a gap.  We present detailed experimental evidences in support of the gap scenario.  This finding reveals a fundamental intrinsic property of epitaxial graphene and demonstrates the possibility of engineering the band gap in epitaxial graphene.

\end{abstract}

\maketitle

With the continuing demand for even smaller transistors, we will soon reach the physical limits of silicon as a host material.  This has motivated the search for novel materials to replace silicon in the past decade.  Among the various candidates, carbon-based materials and in particular carbon nanotubes have shown impressive potential for nanoelectronics applications \cite{CNTreview}.
Most recently attention has turned to planar graphene sheets as an even better alternative thanks to their remarkable electronic transport properties at room temperature \cite{NovoselovSci, NovoselovNat, ZhangNat, BergerSci, RTQHE}, such as an electron mobility that is even higher than that of silicon.  Although the possibility of using graphene to make transistors has been recently explored both in epitaxially grown graphene \cite{BergerSci} and in graphene ribbons  \cite{GeimNatureMat, NetoRMP, Leime}, the biggest hurdle to overcome is that graphene is a zero gap semiconductor.  Over the past few years, different ways to induce a gap in the electronic spectra and generate a finite mass for the Dirac fermions have been proposed.  These range from engineering graphene nanoribbons \cite{KimGap, Nilsson} or quantum dots \cite{Trauzettel, Fertig}, where the gap is the result of the quantum confinement of the electrons in these structures \cite{DresselhausGap}, to applying a voltage to bilayer graphene, i.e. by making the top and bottom layers inequivalent \cite{McCann, Gapadded1, Gapadded2, OhtaSci}, or through the growth of graphene on a substrate which will break the carbon sublattice symmetry \cite{Giovannetti}.

Here we present a detailed study of the electronic structure of single layer epitaxial graphene with the main focus on the $\pi$ bands by using angle resolved photoemission spectroscopy (ARPES).  The dispersion of the $\pi$ bands shows overall agreement with conical dispersions of Dirac fermions.  However, clear deviations from this conical dispersion are observed near the Dirac point.  We examine the two possible explanations for the experimental data: many-body interactions and the opening of a gap at the Dirac point.  Our data suggests that the gap opening is a more likely scenario.

\begin{figure}
\includegraphics[width=8.8cm] {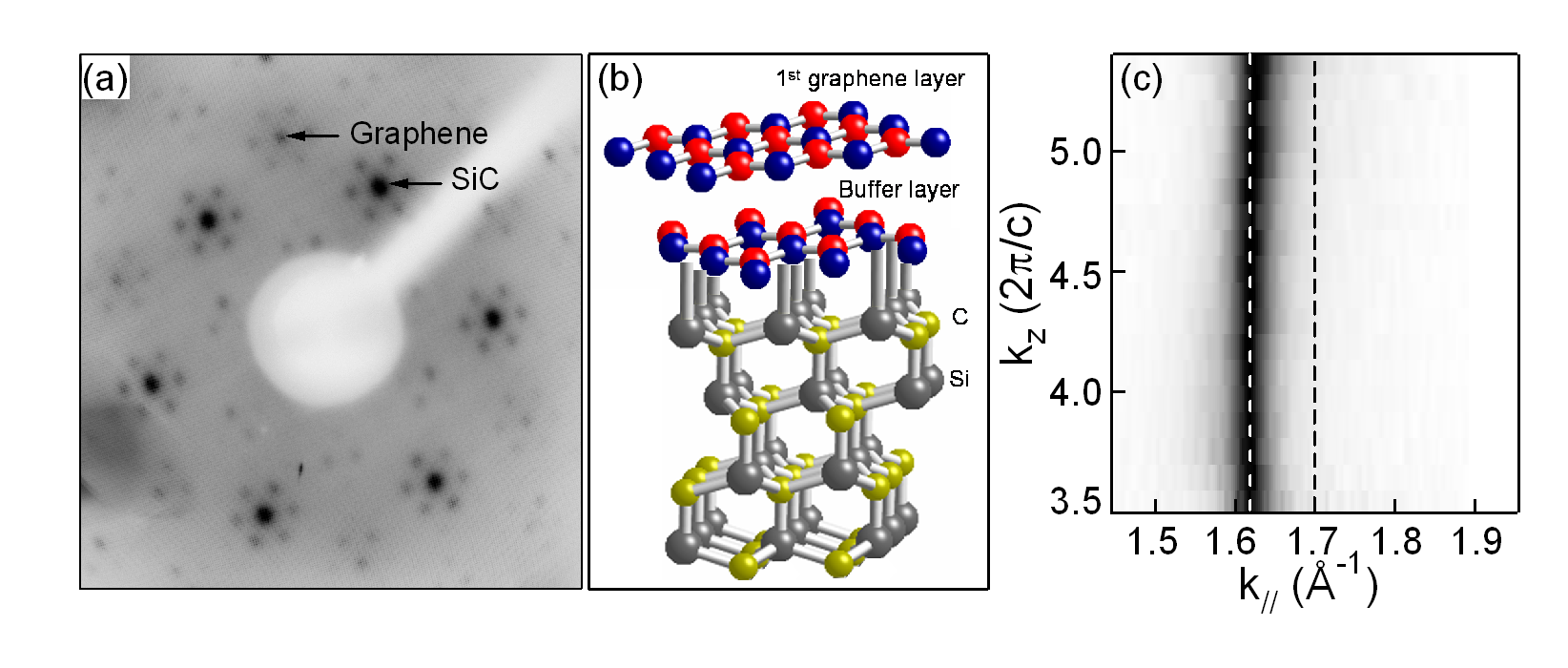}
\label{Figure 1} \caption{(a) LEED pattern of single layer epitaxial graphene on SiC taken at 180eV.  (b) Schematic drawing of the structure of single layer graphene on top of the buffer layer and the SiC substrate.  (c) Intensity map as a function of in-plane and out-of-plane momenta (\kpar and \kz) at -1 eV.}
\end{figure}

The monolayer graphene film was grown epitaxially on the (0001) surface of an n-type 6H-SiC by thermal decomposition of SiC, as previously reported \cite{Forbeaux, BergerJPC, Liz, OhtaSci}.  The surface oxide of SiC is removed either by annealing {\it in situ} at $\approx$ 900C under silicon flux or by hydrogen etching before being loaded into the ultra high vacuum chamber.  The substrate is then heated by electron beam to 1400C and monitored by low-energy electron diffraction (LEED) until graphene is detected.  The substrate preparation and annealing conditions are crucial knobs for determining the quality of the sample as well as its thickness \cite{BergerJPC, OhtaSci, ZhouNatMat}.  We have measured several samples that have been grown under different conditions.  Although the width of ARPES peaks can be affected by sample quality, the effects discussed in this paper are always present in different samples.  The sample thickness was determined by Auger, x-ray photoemission spectroscopy (XPS) and angle resolved photoemission spectroscopy (ARPES), which consistently showed one monolayer thickness for the samples studied here.  ARPES measurements were carried out at Beamline 12.0.1 of the Advanced Light Source (ALS) in Lawrence Berkeley National Lab with an energy resolution better than 25 meV. The samples were measured at 25K.

Fig.1(a) shows a typical LEED pattern of the as-grown monolayer graphene at 180eV electron energy.  The 1$\times$1 spots for SiC are clearly observed together with the $(6\sqrt{3}\times6\sqrt{3})R30$ reconstruction pattern due to multiple diffraction between the SiC substrate and the monocrystalline graphene overlayer. The $(6\sqrt{3}\times6\sqrt{3})R30$ is an indication of the presence of the carbon rich layer that develops before graphene can be grown \cite{FirstBL}, see Fig.1(b). This carbon rich layer, also known as the `buffer layer', shows the same $\sigma$ bands as graphene \cite{Seyller}, suggesting that these two have the same lattice structure.  However, the $\pi$ bands are absent because the $\pi$ orbitals have hybridized with the dangling bonds from the substrate \cite{BLabinitio}. The buffer layer is electronically inactive near E$_F$ and thus one single layer of graphene plus a buffer layer is necessary to give rise to the $\pi$ band dispersions.  In addition to Auger and XPS measurements, we have determined the number of layers by measuring the band structure with ARPES by measuring the dispersions along \kz.  This is a very effective method as the resulting number of quantum states are directly associated with the number of graphene layers \cite{OhtaSci, OhtaPRL, ZhouNatMat}.  The absence of $\pi$ band dispersion along the \kz~direction in two Brillouin zones (BZs) is a strong indication that the sample studied here is a single layer graphene.  

\begin{figure}
\includegraphics[width=8.8cm] {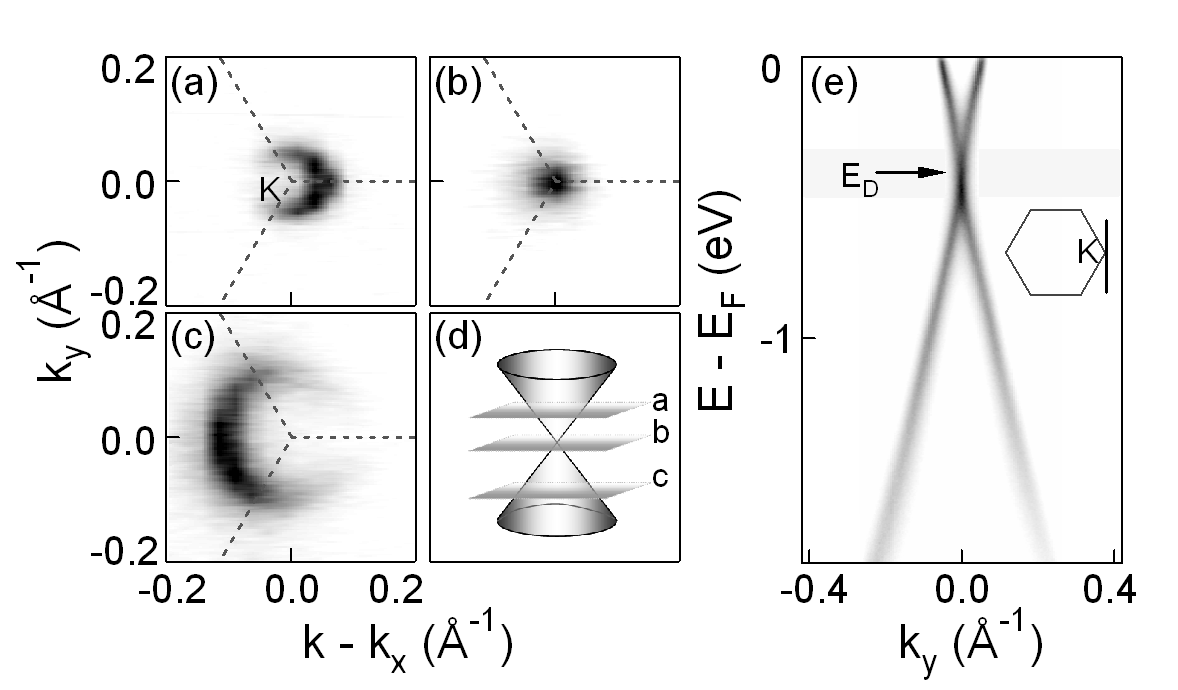}
\label{Figure 2} \caption{(a-c) ARPES intensity maps taken at E$_F$, -0.4 eV and -1.2 eV respectively on single layer graphene.  The dotted line shows the Brillouin zone of graphene. (d) schematic drawing of the dispersion in single layer graphene and the relative energies for data shown in panels a-c. (e) Dispersion of single layer graphene measured along a high symmetric direction through the K point (see black line in the inset).}
\end{figure}

Figs.2(a-c) show ARPES intensity maps at constant energies of E$_F$, -0.4 eV and -1.2 eV.  At E$_F$ (Fig.2(a)), the intensity map shows a finite and almost circular contour centered around the K point. As the binding energy increases, the contour first decreases in size and becomes a point at -0.4 eV (Fig.2(b)).  Beyond -0.4 eV, the size of the contour expands again.  This behavior is consistent with the presence of Dirac fermions, where a conical dispersion centered at the K point is expected.  The Dirac point energy is shifted to 0.4 eV below E$_F$, which shows that the as-grown graphene is electron-doped \cite{OhtaPRL, ZhouNatMat}.  At higher binding energy (Fig.2(c)) the high intensity region in the intensity map deviates from the circular shape.  Similar trigonal distortions have been reported for graphite \cite{ZhouNatPhys}.  Note that the intensity along the circular contour is not isotropic and is strongly suppressed on the left (first BZ) for energy above the Dirac point energy (Fig.2(a)) and on the right (second BZ) for energy below the Dirac point energy (Fig.2(c)). This is a well known effect in graphite and is due to the ARPES dipole matrix element which suppresses or enhances the intensity in different BZs \cite{Shirley}.

Fig.2(e) shows the overall dispersion of the $\pi$ bands measured for a symmetric cut through the K point.  Following the maximum in the intensity plot, two cones dispersing in opposite directions (one upward and another downward) can be clearly distinguished, in overall agreement with the presence of Dirac fermions.  The Dirac point energy, defined as the midpoint between the valence and conduction bands at the K point, occurs at -0.4 eV since the sample is electron doped, as already discussed above.  More importantly, in addition to this shift of the chemical potential and hence E$_D$, deviations from a conical dispersion are observed in the vicinity of the Dirac point.  The first, hardly visible on the energy scale of this figure, is at approximately -0.2 eV below E$_F$ and is due to coupling to phonons \cite{AnnalsPhys, Jessica, Kim}. The second occurs near E$_D$ and is the main focus of this paper.  In particular, we observe that the valence band and the conduction band do not merge at a single point at E$_D$, but instead the top of the conduction bands and the bottom of the valence bands stop before E$_F$ and at the K point there is finite intensity over an extended energy region between these two bands (see shaded area in the figure). 

\begin{figure*}
\includegraphics[width=14.8cm] {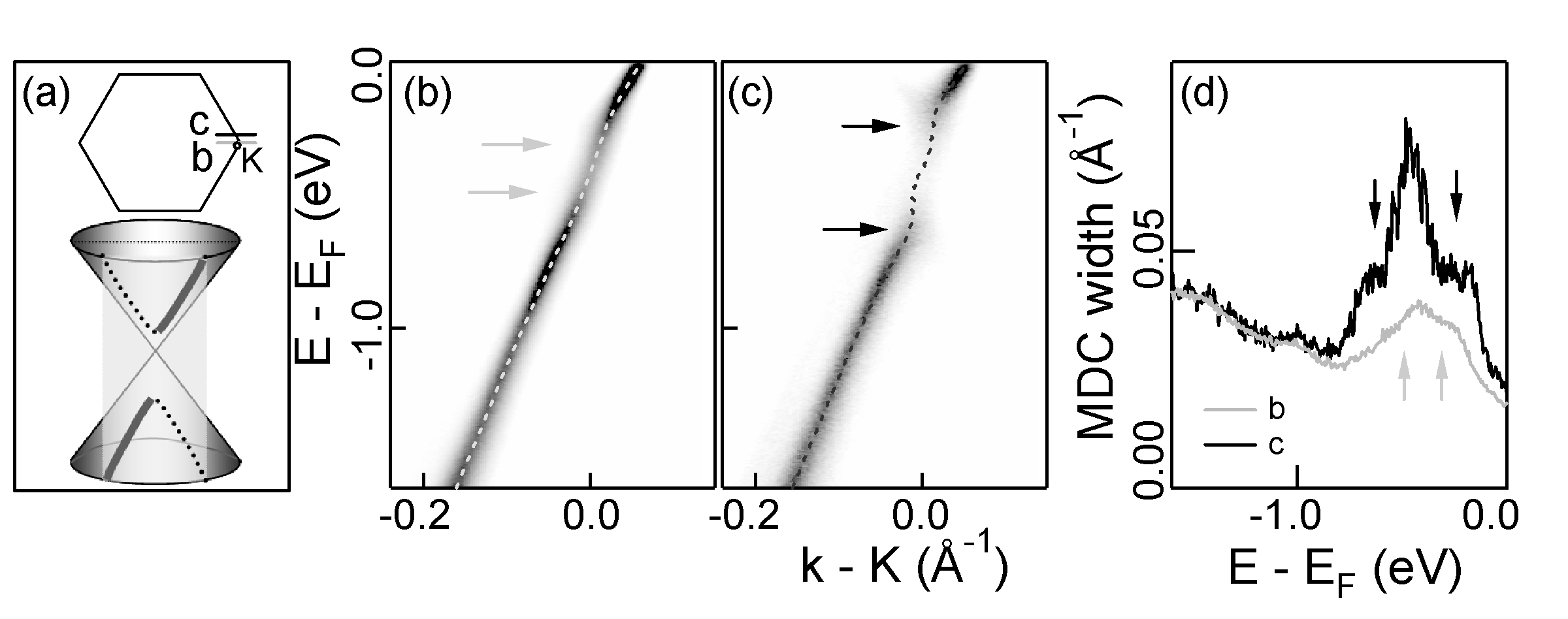}
\label{Figure 3} \caption{(a) Schematic drawing (not to scale) for the cuts shown in (b) and (c) in the BZ of graphene and the conical dispersion. (b,c) Dispersions measured through and off the K point.  The dotted white and dark gray lines are dispersions extracted from the MDCs.  (d) MDC width as a function of energy.}
\end{figure*}

To gain further insights into the nature of the observed anomaly, it is important to perform a quantitative analysis of the data.  This is usually done by extracting the peak positions and the peak width by fitting the momentum distribution curves (MDCs) at fixed energy and the energy distribution curves (EDCs) at fixed momentum.  By fitting the MDCs using Lorentzian functions, the dispersion can be extracted from the peak positions and the scattering rate, a quantity proportional to the imaginary part of the self energy Im${\Sigma}$, can be obtained from the peak width.  However, one of the caveats of the MDC analysis is that it is based on the assumption that whenever there is a peak in an MDC there are electronic states. 

In Fig. 3, we present data taken along two lines (panel a): one through the K point (panel b), and another one parallel to it but far from the K point (panel c).  These particular cuts are convenient because the intensity is strongly suppressed on one side of the K point, thus allowing the MDC to be fitted with single peaks.  This allows for a more reliable fit of the data. The cut far away from the K point is considered because it definitely has a gap, due to the conical nature of the dispersion. Performing an MDC analysis first on the cut through the K point, the anomalous region of Fig.2 is manifested through a kink in the dispersion and a sudden increase of the scattering rate (pointed to by arrows in panels c and d).  Such anomalies might be considered to be due to self-energy effects.  An appealing explanation is that a decay through plasmon emission is responsible for the deviation from the conical dispersion, as recently proposed \cite{EliNaturePhys, DasSarma, MacDonald}.  However, performing a similar analysis on the cut far away from the K point reveals similar features.  In both cases, we can identify a region between the conduction and valence bands where the intensity decreases (see horizontal arrows), the dispersion deviates from the linear behavior (see dashed line in panel b-c), and the scattering rate shows a sudden increase (see panel d). These striking similarities cast doubt on the validity of the many-body interaction scenario \cite{EliNaturePhys}, as this should be able to account for very similar features far away from the K point where a gap is certainly present and is the natural explanation of the data.  We therefore propose that these similarities are simply a manifestation of a gap opening at the Dirac point.
 
\begin{figure}
\includegraphics[width=8.8cm] {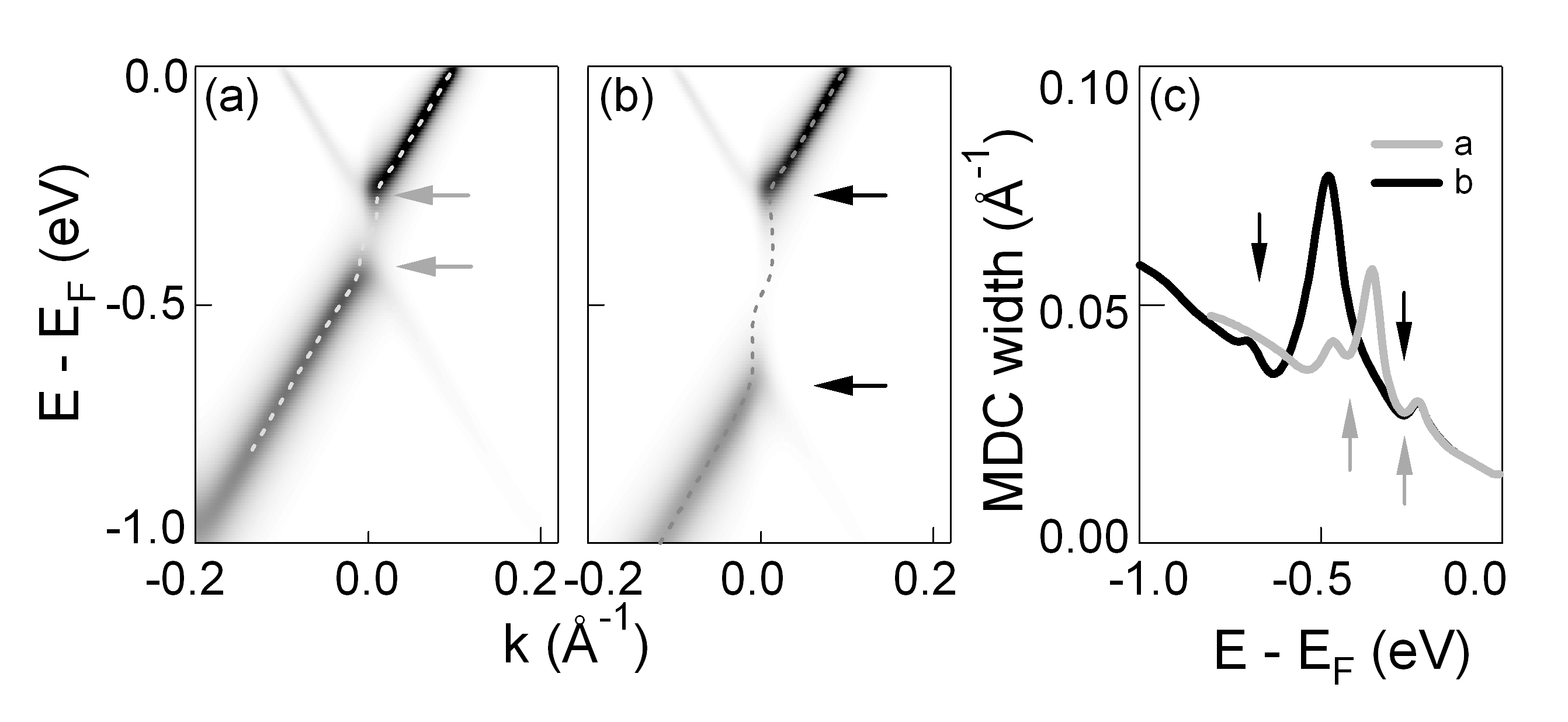}
\label{Figure 4} \caption{(a,b) Simulation of conical dispersions with a gap of 150 and 400 meV.  (c) Extracted MDC width as a function of energy for data shown in panels a and b.}
\end{figure}

It is therefore misleading to discuss the dispersion and scattering rate in the gap region and in general to perform an MDC fit near the bottom or top of a band, despite the possibility that MDC peaks might be present inside the gap region (panel c).  Such peaks exist in our data due to the finite width of the EDCs at the bottom of the conduction band and the top of the valence band. This can be better illustrated in the simulated data shown in Fig.4.  Dirac-like dispersions with finite band gaps of 150 meV (Fig.4(a)) and 400 meV (Fig.4(b)) and hence finite effective masses are used.  The MDC width is modeled to scale linearly with energy with finite constant broadening to account for the finite MDC width at E$_F$.  Matrix elements are included to suppress one side of the dispersion by multiplying the results by a step function with a finite width.  Ignoring the fact that there is a gap between the valence and conduction bands and performing an MDC analysis in this region, the fit shows a kink in the dispersion (Fig.4(a, b)) and an increase in MDC width (Fig.4(c)).  The anomalous region where the MDC width increases scales with the gap region.  This simulation shows that disregarding the absence of electronic states and simply fitting MDCs can sometimes produce misleading results.

\begin{figure}
\includegraphics[width=8.8cm] {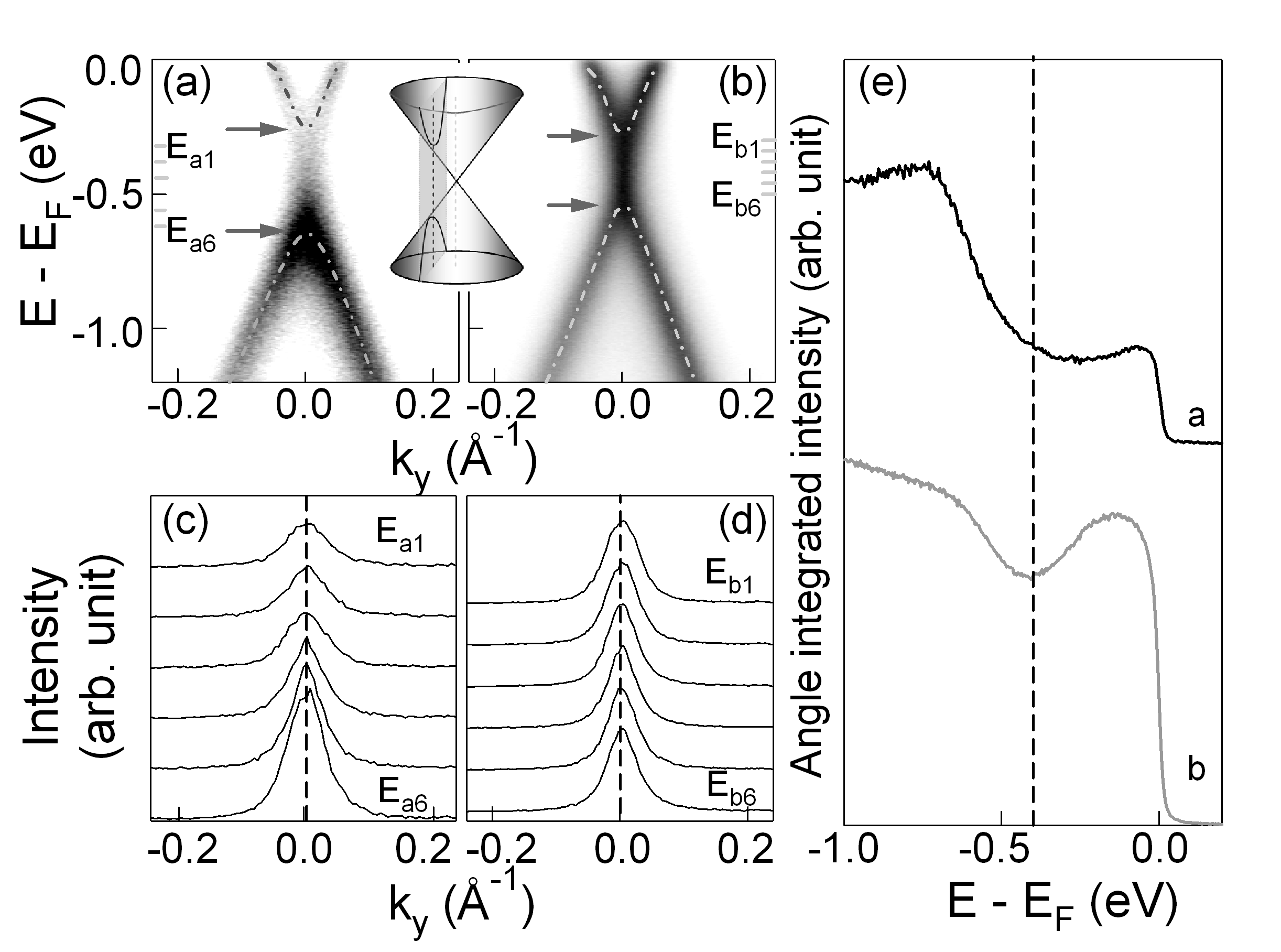}
\label{Figure 5} \caption{(a,b) Dispersions taken along a symmetric direction for cuts away and through the K point.  (c,d) MDCs in the energies as labeled in Fig.5(a,b).  (e) Angle integrated intensity as a function of energy for data in panel a (black curve) and panel b (gray curve).}
\end{figure}

It is clear that in order to extract information on both the conical dispersion and the presence of a gap, one should use a different method.  In Fig.5 we show two cuts, one through the K point and one far away from the K point.  In contrast with Fig 3, the cuts in Fig 5 are now taken symmetrically with respect to the matrix elements.  These cuts can be used to measure both sides of the cones and extract dispersions without artifacts.  The use of this geometry is also important because it is possible to resolve two peaks in the energy distribution curve (EDC) through the K point \cite{ZhouNatMat}, a clear signature of a gap.  

The dispersions extracted from the EDCs are also shown in Fig.5 (dotted lines in Figs.5(a-b)).  
In both cases we can identify a region of energy where the MDC peaks are non-dispersive (Figs.5(c-d)), which coincides with the regions of vertical intensity in the upper panels.
The resemblance between the two cuts is striking and once again suggests that a similar phenomenon is the cause of the vertical region of intensity around the Dirac point (marked by the two horizontal arrows).  Fig.5(e) shows the integrated intensity for the data of Figs.5(a,b), a quantity proportional to the one dimensional density of states.  Though this one dimensional angle integrated intensity does not correspond to the density of states, which is integrated over two dimensions, it gives some hint whether a gap is present.  At high binding energy the dispersion shows an almost linear behavior while near E$_D$ (indicated by broken line) it decreases.  This shows that near E$_D$ there are fewer states, a result that is in agreement with the gap scenario.  When moving away from the K point, the angle integrated intensity curve (black curve) shows a larger anomalous region, which is in agreement with a larger gap.

In summary, ARPES data taken in single layer epitaxial graphene reveals the opening of a gap at the K point.  A possible explanation for this gap is that it is induced by the breaking of the A-B sublattice symmetry \cite{ZhouNatMat} which leads to the rehybridization of the valence and conduction band states associated with the same Dirac point, resulting in a gap at the K point.  In the future, the possibility of doping graphene to move the Fermi energy inside the gap could eventually make graphene a semiconductor suitable for electronic applications.

\begin{acknowledgments}
We would like to thank Gey-Hong Gweon and Jeff Graf for experimental help and useful discussions, A.K. Geim, D.-H. Lee, F. Guinea and A. Castro Neto for useful discussions.
This work was supported by the Director, Office of Science, Office of Basic Energy Sciences, Division of Materials Sciences and Engineering of the U.S Department of Energy under Contract No.~DEAC03-76SF00098; by the National Science Foundation through Grant No.~DMR03-49361 and Grant No.~DMR04-39768.  The Advanced Light Source is supported by the Director, Office of Science, Office of Basic Energy Sciences, of the U.S. Department of Energy under Contract No.\ DE-AC02-05CH11231.   
\end{acknowledgments}

\end{document}